\documentclass[10pt]{article}

\usepackage{amsmath}
\usepackage{amssymb}

\usepackage{graphicx}

\usepackage{url}

\usepackage[labelfont=bf,labelsep=period,justification=raggedright]{caption}

\makeatletter
\renewcommand{\@biblabel}[1]{\quad#1.}
\makeatother

\usepackage{color}

\date{}

\begin{document}

\begin{flushleft}
{\Large
\textbf{A principal component analysis of 39 scientific impact measures.}
}
\\
Johan Bollen$^{1,\ast}$, 
Herbert Van de Sompel$^{1}$, 
Aric Hagberg$^{2,\star}$, and
Ryan Chute$^{1\star}$\\
\bf{1} Digital Library Research and Prototyping Team, Research Library, Los Alamos National Laboratory, Mail Stop P362, Los Alamos NM 87545, USA
\\
\bf{2} Theoretical Division, Mathematical Modeling and Analysis Group, and Center for Nonlinear Studies, Los Alamos National Laboratory, Los Alamos NM 87545, USA
\\
$\ast$ E-mail: Corresponding author: jbollen@lanl.gov, $\star$: Authors made equal contributions.
\end{flushleft}

\begin{tabular}{lp{13cm}}
{\bf Please cite as:} &  Bollen J,  Van de Sompel H,  Hagberg A,  Chute R, 2009 A Principal Component Analysis of 39 Scientific Impact Measures. PLoS ONE 4(6): e6022. doi:10.1371/journal.pone.0006022 \\
URL	&	\texttt{http://www.plosone.org/article/info\%3Adoi\%2F10.1371\%2Fjournal.pone.0006022}\\
\end{tabular}

\section*{Abstract}

\textbf{Background}: The  impact of scientific publications has traditionally been expressed in terms of citation counts.
However, scientific activity has moved online over the past decade. To better capture scientific impact in the digital era,
a variety of new impact measures has been proposed on the basis of social network analysis and usage log data.
Here we investigate how these new measures relate to each other, and how accurately and completely they express
scientific impact. 

\textbf{Methodology}: We performed a principal component analysis of the rankings produced by 39 existing and proposed measures of scholarly impact that were calculated on the basis of both citation and usage log data.

\textbf{Conclusions}: Our results indicate that the notion of scientific impact is a multi-dimensional
construct that can not be adequately measured by any single indicator, although
some measures are more suitable than others. The commonly used citation Impact Factor is not
positioned at the core of this construct, but at its periphery, and should thus be used with caution.


\section*{Introduction}

Science is a \emph{gift economy}; value is defined as the degree to which one's ideas have freely contributed to knowledge and impacted the thinking of others. Since authors use citations to indicate which publications influenced their work, scientific impact can be measured as a function of the citations that a publication receives. Looking for quantitative measures of scientific impact, administrators and policy makers have thus often turned to citation data.\\

A variety of impact measures can be derived from raw citation data. It is however highly common to assess scientific impact in terms of average journal citation rates. In particular, the Thomson Scientific Journal Impact Factor (JIF) \cite{impactreivew:garfield1999} which is published yearly as part of the Journal Citation Reports (JCR) is based on this very principle; it is calculated by dividing the total number of citations that a journal receives over a period of 2 years by the number of articles it published in that same period.\\

The JIF has achieved a dominant position among measures of scientific impact for two reasons. First, it is published as part of a well-known, commonly available citation database (Thomson Scientific's JCR). Second, it has a simple and intuitive definition. The JIF is now commonly used to measure the impact of journals and by extension the impact of the articles they have published, and by even further extension the authors of these articles, their departments, their universities and even entire countries. However, the JIF has a number of undesirable properties which have been extensively discussed in the literature \cite{sensen:opthof1997,impact:seglen1997,impact:harter1997,advant:bordons2002,impact:plos2006}. This had led to a situation in which most experts agree that the JIF is a far from perfect measure of scientific impact  but it is still generally used because of the lack of accepted alternatives.\\

The shortcomings of the JIF as a simple citation statistic have led to the introduction of other measures of scientific impact. Modifications of the JIF have been proposed to cover longer periods of time \cite{mathem:egghe1988} and shorter periods of times (JCR's Citation Immediacy Index). Different distribution statistics have been proposed, e.g.~Rousseau (2005) \cite{median:rousseau2005} and the JCR Citation Half-life\\(\texttt{http://scientific.thomson.com/free/essays/citationanalysis/citationrates/}).\\ The H-index \cite{indexq:hirsch2005} was originally proposed to rank authors according to their rank-ordered citation distributions, but was extended to journals by Braun (2005) \cite{hirsch:braun2005}. Randar (2007) \cite{randar:bihui2007} and  Egghe (2006) \cite{theory:egghe2006} propose the g-index as a modification of the H-index.\\

In addition, the success of Google's method of ranking web pages has inspired numerous measures of journal impact that apply social network analysis \cite{social:wasserman1994} to citation networks. Pinski (1975) \cite{citati:pinski1976} first proposed to rank journals according to their eigenvector centrality in a citation network. Bollen (2006) \cite{journa:bollen2006} and Dellavalle (2007) \cite{refini:dellavalle2007} proposed to rank journals according to their citation PageRank (an approximation of Pinski's eigenvector centrality), followed by the launch of \texttt{eigenfactor.org} that started publishing journal PageRank rankings in 2006. The Scimago group (\texttt{http://www.scimagojr.com/}) now publishes the Scimago Journal Rank (SJR) that ranks journals based on a principle similar to that used to calculate citation PageRank. PageRank has also been proposed to rank individual articles \cite{findin:chen2007}. Using another social network measure, Leydesdorff (2007) \cite{betwee:leydesdorff2007} proposes betweenness centrality as an indicator of a journal's interdisciplinary power.\\

Since scientific literature is now mostly published and accessed online, a number of initiatives have attempted to measure scientific impact from {\it usage log data}. The web portals of scientific publishers, aggregator services and institutional library services now consistently record usage at a scale that exceeds the total number of citations in existence. In fact, Elsevier announced 1 billion fulltext downloads in 2006, compared to approximately 600 million citations in the entire Web of Science database.  The resulting usage data allows scientific activity to be observed immediately upon publication, rather than to wait for citations to emerge in the published literature and to be included in citation databases such as the JCR; a process that with average publication delays can easily take several years. Shepherd (2007) \cite{feasab:shepherd2007} and Bollen (2008)\cite{usage:bollen2008} propose a Usage Impact Factor which consists of average usage rates for the articles published in a journal, similar to the citation-based JIF. Several authors have proposed similar measures based on usage statistics \cite{readin:darmoni2002}. Parallel to the development of social network measures applied to citation networks, Bollen (2005, 2008)\cite{altern:bollen2005,toward:bollen2008} demonstrate the feasibility of a variety of social network measures calculated on the basis of usage networks extracted from the clickstream information contained in usage log data.\\

These developments have led to a plethora of new measures of scientific impact that can be derived from citation or usage log data, and/or rely on distribution statistics or more sophisticated social network analysis. However, which of these measures is most suitable for the measurement of scientific impact? This question is difficult to answer for two reasons. First, impact measures can be calculated for various citation and usage data sets, and it is thus difficult to distinguish the true characteristics of a measure from the peculiarities of the data set from which it was calculated. Second, we do not have a universally accepted, golden standard of impact to calibrate any new measures to. In fact, we do not even have a workable definition of the notion of ``scientific impact'' itself, unless we revert to the tautology of defining it as the number of citations received by a publication. As most abstract concepts  ``scientific impact'' may be understood and measured in many different ways. The issue thus becomes which impact measures best express its various aspects and interpretations.\\

Here we report on a Principal Component Analysis (PCA) \cite{princi:jolliffe2002} of the rankings produced by a total of 39 different, yet plausible measures of scholarly impact. 19 measures were calculated from the 2007 JCR citation data and 16 from the MESUR project's log usage data collection (\texttt{http://www.mesur.org/}). We included 4 measures of impact published by the Scimago (\texttt{http://www.scimagojr.com/}) group that were calculated from Scopus citation data. The resulting PCA shows the major dimensions along which the abstract notion of scientific impact can be understood and how clusters of measures correspond to similar aspects of scientific impact.

\section*{Methods}

The mentioned 39 scientific impact measures were derived from various sources. Our analysis included several existing measures that are published on a yearly basis by Thomson-Reuters and the Scimago project. Other measures were calculated on the basis of existing citation- and usage data. The following sections discuss the methodology by which each of these impact measures was either extracted or derived from various usage and citation sources.

\subsection*{Data preparation and collection}

\begin{figure}[ht!]
\begin{center}
\includegraphics[width=5in]{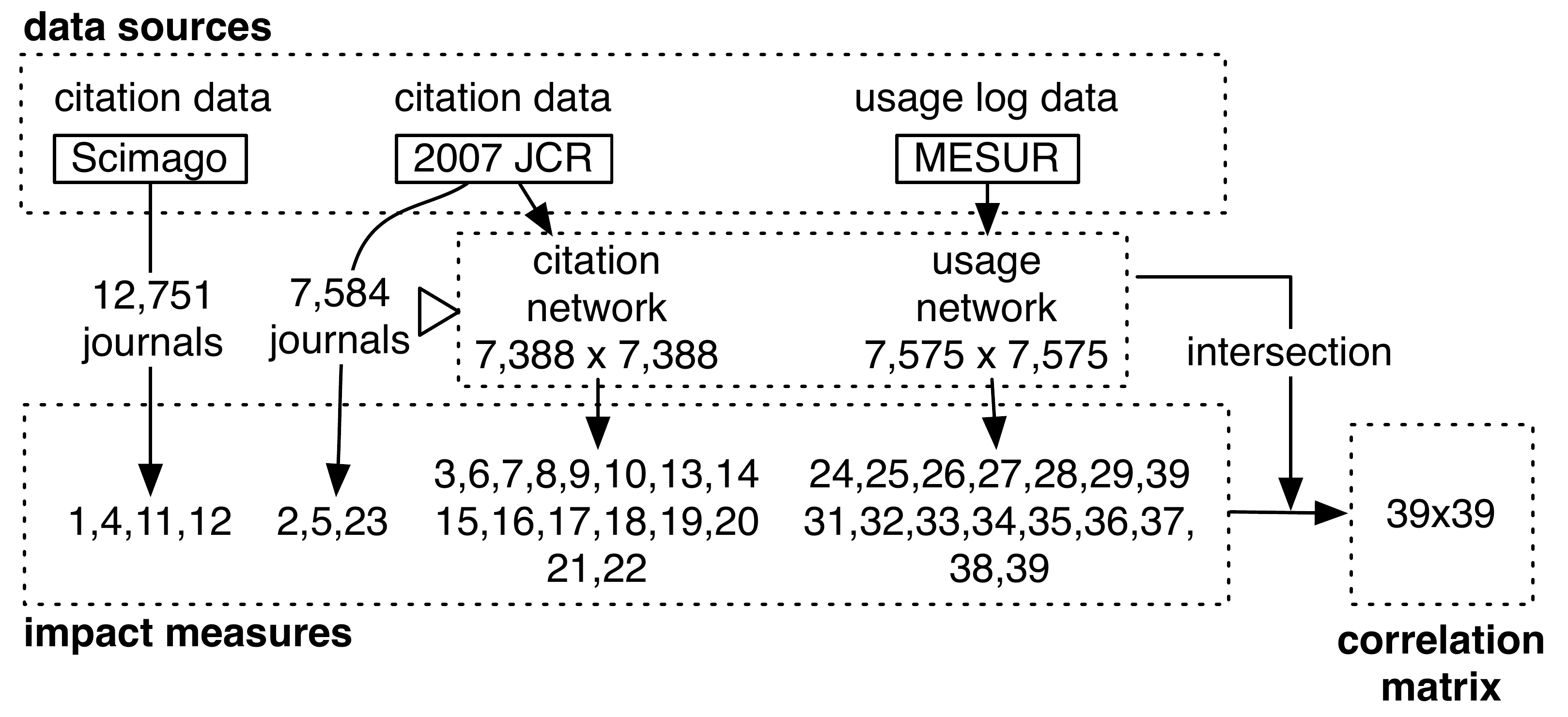}
\caption{\label{dataflow}Schematic representation of data sources and processing. Impact measure identifiers refer to Table \ref{measure_loadings}.}
\end{center}
\end{figure}

As shown in Fig. \ref{dataflow}, the following databases were used in this analysis:
\begin{description}
	\item[Citation] The CDROM version of the 2007 Journal Citation Reports (JCR Science and Social Science Editions) published by Thomson-Reuters Scientific (formerly ISI)
	\item[Usage] The MESUR project's reference collection of usage log data: \texttt{http://www.mesur.org/}: a collection of  346,312,045 user interactions recorded by the web portals operated by Thomson Scientific (Web of Science), Elsevier (Scopus), JSTOR, Ingenta, University of Texas (9 campuses, 6 health institutions), and California State University (23 campuses) between March 1st 2006 and February 1st 2007.
	\item[Additional citation measures] A set of journal rankings published by the Scimago project that are based on Elsevier Scopus citation data:\\ \texttt{http://www.scimagojr.com/}
\end{description}

In the following sections we detail the methodology that was used to retrieve and calculate 39 scientific impact measures from these data sets, and the subsequent analysis of the correlations between the rankings they produced. Throughout the article measures are identified by a unique identifier number that is listed in Table \ref{measure_loadings}. We hope these identifiers will allow readers to more conveniently identify measures in subsequently provided diagrams and tables such as Fig. \ref{dataflow}, \ref{PCA_m39_map} and \ref{hclust_m39}.

\begin{table*}
\begin{small}
\caption{\label{measure_loadings} Measure loadings after varimax rotation of first 2 components of PCA analysis of measure correlations (Spearman rank-order). Average Spearman rank-order correlations to all other measures are listed under $\bar{\rho}$ (five lowest values indicated by $\star$).}
\begin{center}
\begin{tabular}{lllllrrl}
\hline\hline
ID		&	Type		&	Measure					&		Source			&	Network parameters		&	PC1			& 	PC2			&	 $\bar{\rho}$\\\hline\hline
1			&	Citation	&	Scimago Journal Rank	&		Scimago/Scopus	&						&	-0.974		& 	-8.296		&	0.556$^\star$	\\
2			&	Citation	&	Immediacy Index		&		JCR 2007			&						&	1.659		& 	-7.046		&	0.508$^\star$	\\
3			&	Citation	&	Closeness Centrality	& 		JCR 2007			&	Undirected, weighted	&		0.339	& 	-6.284		&	0.565$^\star$	\\
4			&	Citaton	&	Cites per doc			&		Scimago/Scopus	&						& 		-1.311	& 	-6.192		& 	0.588$^\star$	\\
5			&	Citation	&	Journal Impact Factor	&		JCR 2007			&						& 		-1.854	& 	-5.937		&	0.592$^\star$	\\
6			&	Citation	&	Closeness centrality		&		JCR 2007			&	Undirected, unweighted	&		-1.388	& 	-4.827		&	0.619		\\
7			&	Citation	&	Out-degree centrality	&		JCR 2007			&	Directed, weighted		& 		-3.191	& 	-4.215		&	0.642		\\
8			&	Citation	&	Out-degree centrality	&		JCR 2007			&	Directed, unweighted	&		-2.703	& 	-4.015		&	0.640		\\
9			&	Citation	&	Degree Centrality		& 		JCR 2007			&	Undirected, weighted	&		-4.850	& 	-2.834		&	0.690		\\
10			&	Citation	&	Degree Centrality		&		JCR 2007			&	Undirected, unweighted	&		-4.398	& 	-2.643		&	0.691		\\
11			&	Citation	&	H-Index				&		Scimago/Scopus	&						& 		-3.326	& 	-2.003		&	0.681		\\
12			&	Citation	&	Scimago Total cites		&		Scimago/Scopus	&						& 		-4.926	& 	-1.722		&	0.712		\\
13			&	Citation	&	Journal Cite Probability	&		JCR 2007			&						&		-5.389	& 	-1.647		&	0.710		\\
14			&	Citation	&	In-degree centrality		&		JCR 2007			&	Directed, unweighted	&		-5.302	& 	-1.429		&	0.717		\\
15			&	Citation	&	In-degree	centrality		&		JCR 2007			&	Directed, weighted		& 		-5.380	& 	-1.554		&	0.712		\\
16			&	Citation	&	PageRank			& 		JCR 2007			&	Directed, unweighted	&		-4.476	& 	0.108		&	0.693		\\
17			&	Citation	&	PageRank			&		JCR 2007			&	Undirected, unweighted	&		-4.929	& 	0.731		&	0.726		\\
18			&	Citation	&	PageRank			&		JCR 2007			&	Undirected, weighted	&		-4.160	& 	0.864		&	0.696		\\
19			&	Citation	&	PageRank			&		JCR 2007			&	Directed, weighted		& 		-3.103	& 	0.333		&	0.659		\\
20			&	Citation	&	Y-factor				&		JCR 2007			&	Directed, weighted		& 		-2.971	& 	0.317		&	0.657		\\
21			&	Citation	&	Betweenness centrality	&		JCR 2007			&	Undirected, weighted 	& 		-0.462	& 	0.872		&	0.643		\\
22			&	Citation	&	Betweenness centrality	&		JCR 2007			&	Undirected, unweighted	& 		-0.474	& 	1.609		&	0.642		\\
23			& {\it Citation}	&	{\it Citation Half-Life}		&		{\it JCR 2007}		&						&		/		&	/			&	{\it 0.037}		\\\
24			&	Usage	&	Closeness centrality		& 		MESUR 2007		&	Undirected, weighted	&		3.130	& 	2.683		&	0.703		\\
25			&	Usage	&	Closeness centrality		&		MESUR 2007		&	Undirected, unweighted	& 		3.100	& 	3.899		&	0.731		\\
26			&	Usage	&	Degree centrality		&		MESUR 2007		&	Undirected, unweighted	& 		3.271	& 	3.873		&	0.729		\\
27			&	Usage	&	PageRank			&		MESUR 2007		&	Undirected, unweighted	& 		3.327	& 	4.192		&	0.728		\\
28			&	Usage	&	PageRank			&		MESUR 2007		&	Directed, unweighted	& 		3.463	& 	4.336		&	0.727		\\
29			&	Usage	&	In-degree centrality		&		MESUR 2007		&	Directed, unweighted	& 		3.463	& 	4.015		&	0.728		\\
30			&	Usage	&	Out-degree centrality	&		MESUR 2007		&	Directed, unweighted	& 		3.484	& 	3.994		&	0.727		\\
31			&	Usage	&	PageRank			&		MESUR 2007		&	Directed, weighted		& 		3.780	& 	4.217		&	0.710		\\
32			&	Usage	&	PageRank			&		MESUR 2007		&	Undirected, weighted	& 		3.813	& 	4.223		&	0.710		\\
33			&	Usage	&	Betweenness centrality	&		MESUR 2007		&	Undirected, unweighted	& 		3.988	& 	4.271		&	0.699		\\
34			&	Usage	&	Betweenness centrality	&		MESUR 2007		&	Undirected, weighted 	& 		3.957	& 	3.698		&	0.693		\\
35			&	Usage	&	Degree centrality		&		MESUR 2007		&	Undirected, weighted	& 		5.293	& 	3.528		&	0.683		\\
36			&	Usage	&	Out-degree centrality	&		MESUR 2007		&	Directed, weighted		&  		5.302	& 	3.518		&	0.683		\\
37			&	Usage	&	In-degree	centrality		&		MESUR 2007		&	Directed, weighted		& 		5.286	& 	3.531		&	0.683		\\
38			&	Usage	&	Journal Use Probability	&		MESUR 2007		&	 					&		8.914	& 	1.833		&	0.593		\\
39			& {\it Usage}	&	{\it Usage Impact Factor}	&		{\it MESUR 2007}	&						&		/		&	/			&	{\it 0.279}		\\
\hline\hline
\end{tabular}
\end{center}
\end{small}
\end{table*}

\subsection*{Retrieving existing measures}

The 2007 JCR contains a table listing 4 citation-based impact measures for a set of approximately 7,500 selected journals, namely

\begin{description}
	\item[2007 Immediacy Index] (Table \ref{measure_loadings}, ID 2):The same year average citation rate, i.e.~the average number of times articles that were published in a journal in 2006 were cited in 2006.
	\item[2007 Journal Impact Factor]  (Table \ref{measure_loadings}, ID 5): A 2 year average per-article citation rate of a journal, i.e. the average number of times articles that were published in a journal in 2004 and 2005 were cited in 2006.
	\item[Citation Half-life] (Table \ref{measure_loadings}, ID 23):The median age of articles cited in a journal in 2006.
\end{description}

\vspace{0,5cm}

In addition, the Scimago project publishes several impact measures that are based on Elsevier's Scopus citation data. We retrieved the following 4 measures from its web site:

\begin{description}
	\item[2007 Scimago Journal Rank] (Table \ref{measure_loadings}, ID 1) The citation PageRank of a journal calculated on the basis of Elsevier Scopus citation data divided by the number of articles published by the journal in the citation period (3 years) (\texttt{http://www.scimagojr.com/SCImagoJournalRank.pdf}), i.e.~an average per-article journal PageRank.
	\item[Cites per doc] (Table \ref{measure_loadings}, ID 4) The average number of citations received by articles published in a year over a 2 year period in the Scopus database.
	\item[H-Index] (Table \ref{measure_loadings}, ID 11) Journal citation h-index, i.e.~the $h$ number of articles in a journal that received at least $h$ citations \cite{hirsch:braun2005} in the Scopus database.
	\item[Scimago Total cites] (Table \ref{measure_loadings}, ID 12) The number of citations received by the articles published in a journal during the three previous years according to the Scopus database.
\end{description}

The Scimago journal rankings were downloaded from their web site in the form of an Excel spreadsheet and loaded into a MySQL database. This added 4 measures of journal impact to our data set bringing the total number of retrieved, existing measures to 8.

\subsection*{Calculating social network measures of scientific impact.}

In \cite{altern:bollen2005}  and \cite{journa:bollen2006} we describe methods to rank journals on the basis of various social network measures of centrality \cite{social:wasserman1994}, e.g.~betweenness centrality that is calculated from journal citation- and usage graphs. These social network measures were shown to elucidate various aspects of a journal's scientific impact on the basis of its connections in citation- or usage-derived networks. In addition, this approach has led to innovative ranking services such as \url{eigenfactor.org}. We followed the same approach in this work by extracting citation- and usage-networks from our data and defining a set of well-studied social network measures on the basis of those networks as shown in Fig. \ref{dataflow}. In the following sections, we therefore first describe the creation of the citation- and usage networks after which we describe the set of social network measures that were calculated on the basis of both.

\subsubsection*{Citation network}

The 2007 JCR contains a table that lists the number of citations that point from one journal to another. The number of citations is separated according to the publication year of both the origin and target of the citation.
For example, from this table we could infer that 20 citations point from articles published in "Physica Review A" in 2006 to articles published in "Physica Review B" in 2004 and 2005. Each such data data point can thus be described as the n-tuple 

\[a \in A=V^2 \times Y_s \times Y_e \times \mathbb{N}^+\]

where $V=\{v_1, \cdots, v_n\}$  is the set of $n$ journals for which we have recorded citation data, $Y_s=\{y_0, \cdots, y_m\}$ is the set of $m$ years for which outgoing were recorded, $Y_e=\{y_0, \cdots, y_k\}$ is the set of $k$ years for which incoming citations were recorded, and $\mathbb{N}^+$ denotes the set of positive integers including zero that represent the number of counted citations. For example, the journal citation tuplet $a=(1, 2, \{2006\}, \{2004,2005\}, 50)$ represents the observation that 20 citations point from articles published in journal 1 in the year 2006 to those published in journal 2 in 2004 and 2005.\\

$A$, the set of citation n-tuples, describes a citation network whose connections indicate the number of times that articles published in one journal cited the articles published in another journal for a particular time period. Such a network can be represented by the citation matrix $C_{Y_s,Y_e}$ of which each entry $c_{i,j}$ represents the number of observed citations that point from articles published in journal $v_i$ in the date range given by $Y_s$ to articles published in journal $v_j$ in the date range $Y_e$.\\

We attempted to ensure that our citation network conformed to the definition of the Journal Impact Factor rankings published in the 2007 JCR. We therefore extracted citations from the JCR that originated in 2006 publications and pointed to 2004 and 2005 publications. The resulting citation network contained 897,608 connections between 7,388 journals, resulting in a network density of 1.6\% (ratio of non-zero connections over all possible non-reflexive connections). This citation network was represented as a $7,338 \times 7,338$ matrix labeled C whose entries $c_{i,j}$ were the number of 2006 citations pointing from journal $i$ to the 2004 and 2005 articles of journal $j$.

\subsubsection*{Usage network}

In \cite{clicks:bollen2009} we describe a methodology to derive journal relations from the session clickstreams in large-scale usage data. The same methodology has in this case been used to create a journal usage network on the basis of which a set of social network measures were calculated. This procedure, related to association rule learning \cite{mining:aggarwal1998}, is described in more detail in Bollen (2006, 2008) \cite{altern:bollen2005,toward:bollen2008} with respect to the calculation of usage-based, social-network measures of scientific impact.\\

In short, the MESUR project's reference collection of usage log data consists of log files recorded by a variety of scholarly web portals (including some of the world's most significant publishers and aggregators) who donated their usage log data to the MESUR project in the course of 2006-2007. All MESUR usage log data consisted of a list of temporally sorted ``requests''. For each individual request the following data fields were recorded: (1) date/time of the request, (2) session identifier, (3) article identifier, and (4) request type. The session identifier grouped requests issued by the same (anonymous) user, from the same client, within the same session. This allowed the reconstruction of user ``clickstreams'', i.e.~the sequences of  requests by individual users within a session. Since each article for this investigation is assumed to be published in a journal, we can derive journal clickstreams from article clickstreams.\\

Over all clickstreams we can thus determine the transition probability

\[P(i,j) = \frac{N(v_i,v_j)}{\sum_j N(v_i,v_j)}\]

where $N(v_i,v_j)$ denotes the number of times that we observe journal $v_i$ being followed by $v_j$ in the journal clickstreams in MESUR's usage log data. The transition probability $P(i,j)$ thus expresses the probability by which we expect to observe $v_j$ after $v_i$ over all user clickstreams.\\

This analysis was applied to the MESUR reference data set, i.e.~346,312,045 user interactions recorded by the web portals operated by Thomson Scientific (Web of Science), Elsevier (Scopus), JSTOR, Ingenta, University of Texas (9 campuses, 6 health institutions), and California State University (23 campuses) between March 1st 2006 and February 1st 2007. To ensure that all subsequent metrics were calculated over the same set of journals, the resulting set of journal transition probabilities were trimmed to $7,575$ journals for which a JIF could be retrieved from the 2007 JCR. All usage transition probabilities combined thus resulted in the $7,575 \times 7,575$ matrix labeled $U$. Each entry $u_{i,j}$ of matrix $U$ was the transition probability $P(i,j)$ between two journals $i$ and $j$.  Matrix $U$ contained 3,617,368 non-zero connections resulting in a network density of 6.3\%.  This procedure and the resulting usage network is explained in detail in \cite{clicks:bollen2009}.

\subsubsection*{Social network measures}

Four classes of social network measures were applied to both the citation and usage network represented respectively by matrix $C$ and matrix $U$, namely:

\begin{description}
	\item[Degree centrality] (Table \ref{measure_loadings}, IDs 7-10, 14, 15, 26, 29, 30, 35-37) 		Number of connections pointing to or emerging from a journal in the network.
	\item[Closeness centrality] (Table \ref{measure_loadings}, IDs 3, 6, 24, 25)					The average length of the geodesic connecting a specific journal to all other journals in the network.
	\item[Betweenness centrality] (Table \ref{measure_loadings}, IDs 21, 22, 33, 34)				The number of geodesics between all pairs of journals in the network that pass through the specific journal.
	\item[PageRank] (Table \ref{measure_loadings}, IDs 16-19, 27, 28, 31, 32)					As defined by Brin and Page (1998) \cite{anatom:brin1998} and applied to citation networks by Bollen (2006) \cite{journa:bollen2006}.
\end{description}

The definitions of each of the measures in these classes were varied according to the following network factors: (1) Weighted vs. unweighted connections, i.e.~measures can be calculated by assuming that each non-zero connection valued 1 vs.~taken into account the actual weight of the connection, (2) Directed vs. undirected connections, i.e. some measures can be calculated to take into account the directionality of journal relations or not, and finally (3) Citation vs. usage network data, i.e. any of these measure variations can be calculated for either the citation or the usage network.\\

These factors result in $2^3=8$ variations for each the above listed 4 classes of social network measures, i.e. 32 variants. However, not all permutations make equal sense. For example, in the case of Betweenness Centrality we calculated only two of these variants that both ignored connection directionality (irrelevant for betweenness) but one took into account connection weights (weighted geodesics) and another ignored connections weights (all connections weighted $>0$). Each of these variants were however calculated for the citation and usage-network. The final list of social network measures thus to some degree reflect our judgment on which of these permutations were meaningful.

\subsection*{Hybrid Measures}

In addition to the existing measures and the social network measure, we calculated, a number of measures that did not fit any the above outlined classes, namely

\begin{description}
	\item[Y-Factor]  (Table \ref{measure_loadings}, ID 20)	A measure that results from multiplying a journal's Impact Factor with its PageRank, described in Bollen (2006) \cite{journa:bollen2006}.
	\item[Journal Cite Probability] (Table \ref{measure_loadings}, ID 13) We calculated the Journal Cite Probability  from the citation numbers listed in the 2007 JCR 2007.
	\item[Journal Use Probability]  (Table \ref{measure_loadings}, ID 38) The normalized frequency by which a journal will be used according to the MESUR usage log data.
	\item[Usage Impact Factor] (Table \ref{measure_loadings}, ID 39) Same definition as the JIF, but expressing the 2-year "usage" average for articles published in a journal.
\end{description}

\subsection*{Measures overview}

In total, we calculated 32 citation- and usage-based impact measures; 16 social network measures on the basis of matrix $C$ (citation network) and 16 social network measures on the basis of matrix $U$ (usage network). 4 journal impact measures published by the Scimago group (\texttt{http://www.scimagojr.com/}) and 3 pre-calculated impact measures from the 2007 JCR were added, bringing the total to 39 measures. A list of measures is provided in Table \ref{measure_loadings} along with information on the data they have been derived from and the various network factors that were applied in their calculation. A list of mathematical definitions is provided in Appendix S1.\\

The set of selected measures was intended to capture the major classes of statistics and social network measures presently proposed as alternatives to the JIF.  In summary, the set of all measures can be categorized in 4 major classes. First, {\em citation and usage statistics} such as Citation Probability (number of one journal's citations over total citations), Usage Probability (amount of one journal's usage over total usage), the JIF, the Scimago Cites per Doc, and a Usage Impact Factor (UIF) whose definition follows that of the JIF but is based on usage counts. Second,  {\em citation and usage social network measures} such as Closeness Centrality (the mean length of geodesics between a journal and all other journals), Betweenness Centrality (number of times that a journal sits on the geodesics between all pairs of journals) and PageRank (cf. Eigenvector Centrality). Third, a set of {\em citation and usage degree centrality} measures such as Out-Degree Centrality, In-Degree Centrality and Undirected Degree Centrality. Finally, we included a set of recently introduced measures such as the Scimago Journal Rank (SJR), the Y-factor (Bollen, 2007) \cite{journa:bollen2006}, the Scimago H-index and Scimago Total Cites.\\

\subsection*{Analysis}

Spearman rank-order correlations were then calculated for each pair of journal rankings. Because $C$, $U$ and the Scimago rankings pertained to slightly different sets of journals, correlation values were only calculated for the intersections of those sets, i.e.~N=7,388, N=7,575 or N=6,913 journals.  For 39 measures. this resulted in a $39 \times 39$ correlation matrix $R$ of which each entry $r_{i,j} \in [-1,1]$ is the Spearman rank-order correlation between the journal rankings produced by measure $i$ and measure $j$.\\

A sample of matrix R for 10 selected measures is shown below. For example, the Spearman rank-order correlation between the Citation H-index and Usage PageRank is 0.66. The IDs listed in Table \ref{measure_loadings} precede each measure name.\\

\begin{equation*}
R_{10\times10}=\begin{pmatrix} 
1.00		&	0.71		&	0.77		&	0.52 		&	0.79		&	0.55		&	0.69		&	0.63		&	0.60		&	0.18\\
0.71		&	0.99		&	0.52		&	0.69 		&	0.79 		&	0.85		&	0.49		&	0.44		&	0.49		&	0.22\\
0.77		&	0.52		&	1.00 		&	0.62		&	0.63		&	0.39		&	0.70		&	0.73		&	0.68		&	0.20\\
0.52		&	0.69		&	0.62 		&	1.00		&	0.68		&	0.78		&	0.49		&	0.56		&	0.65		&	0.06\\
0.79		&	0.79		&	0.63		&	0.68		&	1.00 		&	0.82		&	0.66		&	0.62		&	0.66		&	0.15\\
0.55		&	0.85		&	0.39		&	0.78		&	0.82		&	1.00		&	0.40		&	0.40		&	0.50		&	0.13\\
0.69		&	0.49		&	0.70		&	0.49 		&	0.66		&	0.40		&	1.00		&	0.89		&	0.85		&	0.53\\
0.63		&	0.44		&	0.73		&	0.56		&	0.62		&	0.40		&	0.89		&	1.00		&	0.97		&	0.45\\
0.60		&	0.49		&	0.68 		&	0.65		&	0.66		&	0.50		&	0.85		&	0.97		&	1.00 		&	0.42\\
0.18		&	0.22 		&	0.20		&	0.06 		&	0.15		&	0.13		&	0.53		&	0.45		&	0.42 		&	1.00
\end{pmatrix}
\begin{array}{l}
\mbox{19: Citation PageRank}\\
\mbox{5: Journal Impact Factor}\\
\mbox{22: Citation Betweenness}\\
\mbox{6: Citation Closeness}\\
\mbox{11: Citation H-index}\\
\mbox{1: Citation Scimago Journal Rank}\\
\mbox{31: Usage PageRank}\\
\mbox{34: Usage Betweenness}\\
\mbox{24: Usage Closeness}\\
\mbox{39: Usage Impact Factor}\\
\end{array}
\end{equation*}

Not all pair-wise correlations were statistically significant. Two measures in particular lacked significant correlations ($N=39, p>0.05$) with any of the other measures, namely Citation Half-Life and the UIF. They were for that reason removed from the list of measures under consideration. All other Spearman rank-order correlations were statistically significant ($U$: $N=39, p<0.05$). The reduced $37 \times 37$ correlation matrix $R$ was subjected to a Principal Component Analysis \cite{princi:jolliffe2002} which by means of an eigenvalue decomposition identified 37 orthogonal components of the original correlation matrix $R$.\\

The resulting PCA components were ranked according to the degree by which they explain the variances in $R$'s values (eigenvalues transformed to component loadings). The component loadings are listed in Table \ref{PC_loadings}. The first component, PC1, represents 66.1\% of the variance in measure correlations, with each successive component representing less variance, i.e.~PC2 17\%, PC3 9\% and PC4 4\%. Retention of the first 2 components will thus yield a model that covers 83.4\% of variance in measure correlations. The addition of the third component will yield a model that covers 92.6\% of variation in measure correlations.\\

\begin{table*}
\caption{\label{PC_loadings} Component loadings of Principal Component Analysis of journal ranking correlations (37 measures).}
\begin{center}
\begin{tabular}{l||lllll}
                         		&	PC1		&	PC2		&	PC3		&	PC4		&	 PC5  		\\\hline\hline
Proportion of Variance	&	66.1\%	&	17.3\%	&	9.2\%	&	4.8\%	&	0.9\%		\\
Cumulative Proportion  	&	66.1\%	&	83.4\%	&	92.6\%	&	97.4\%	&	98.3\%		\\\hline\hline
\end{tabular}
\end{center}
\end{table*}

\begin{figure}[ht!]
\includegraphics[width=6.5in]{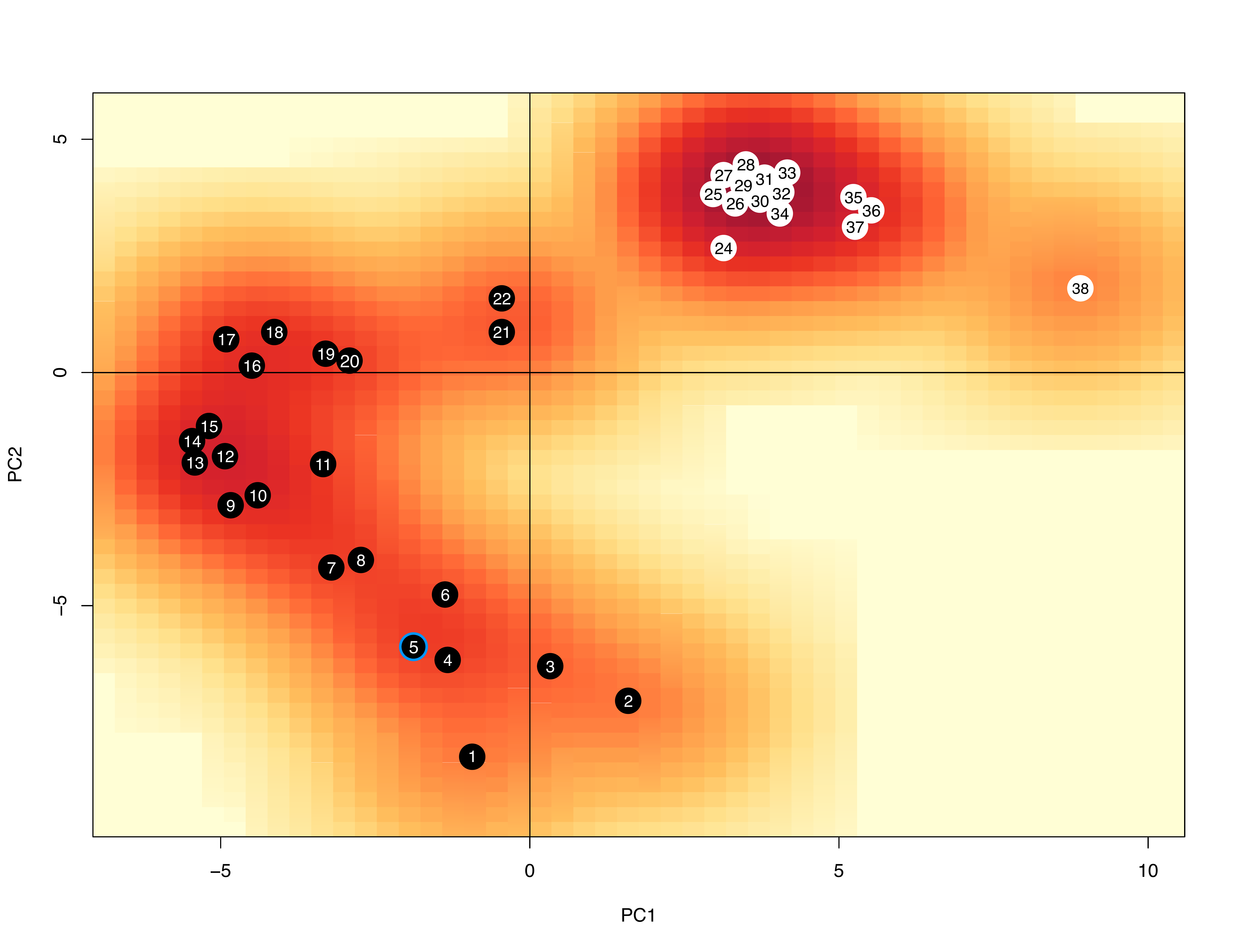}
\caption{\label{PCA_m39_map} Correlations between 37 measures mapped onto first two principal components (cumulative variance=83.4\%) of PCA. Black dots indicate citation-based measures. White dots indicate usage-based measures. The Journal Impact Factor (5) has a blue lining. Measures 23 and 39 excluded.}
\end{figure}

We projected all measures unto the first two components, PC1 and PC2, to create a 2-dimensional map of measures. A varimax rotation was applied to the measure loadings to arrive at a structure that was more amenable to interpretation. The measure loadings for each component are listed in Table \ref{measure_loadings} ("PC1" and "PC2"). The resulting 2-dimensional map of measure similarities is shown in Fig. \ref{PCA_m39_map}. Measures are identified in the map by their ``ID'' in Table \ref{measure_loadings}. Black circles indicate citation-based measures. White circles indicate usage-based measures. The JIF is marked by a blue circle (ID 5). The hue of any map location indicates how strongly measures are concentrated in that particular area, i.e.~red means highly clustered.\\

To cross-validate the PCA results, a hierarchical cluster analysis (single linkage, euclidean distances over $R$'s row vectors) and a k-means cluster analysis were applied to the measure correlations in $R$ to identify clusters of measures that produce similar journal rankings.

\begin{figure}[ht!]
\begin{center}
\includegraphics[width=6.5in]{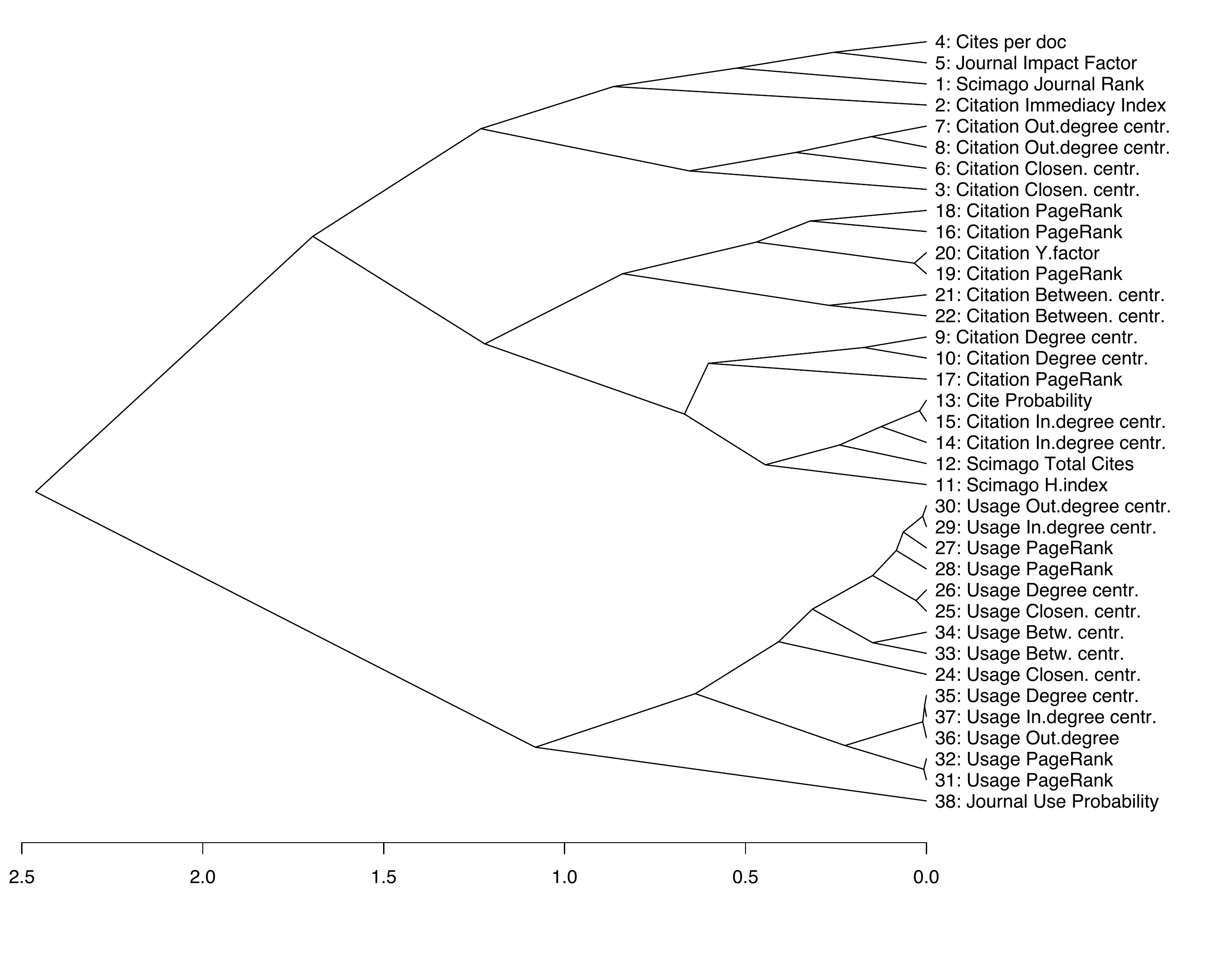}
\caption{\label{hclust_m39} Hierarchical cluster analysis of 39 impact measures (excluding measures 23 and 39).}
\end{center}
\end{figure}

\section*{Results and discussion}

\subsection*{Results}
The map in Fig. \ref{PCA_m39_map} reveals a number of clusters. First, we observe a cluster in the top right quadrant that contains all usage-based measures (IDs 24-37), with the exception of Usage Probability (ID 38). In the upper-left and bottom-left quadrants of the map we find most citation-based measures.  The bottom-left quadrant contains the JIF that is among others surrounded by the Scimago Cites per Doc , the Scimago Journal Rank, the JCR immediacy index (IDs 1-8)  and in the upper section the various permutations of citation degree centrality measures (IDs 9-10, 14-15), a group of Total Cite measures (IDs 12-13) and most prominently the H-index (ID 11). The arrangement of the H-index and Citation Total Cites is quite similar to that found by Leydesdorff (2007) \cite{visual:leydesdorff2007}. The upper-left quadrant nearly uniquely contains citation PageRank and Betweenness Centrality measures (IDs 16-22). The Y-factor (ID 20) is naturally positioned between the two clusters since it is defined as the product of citation PageRank and the JIF.\\

A complete linkage hierarchical cluster analysis based on the Euclidean distances of the measure $R$'s row vectors confirms these general distinctions. When we cut the dendrogram in Fig. \ref{hclust_m39} at the $1.1$ distance level, we find 4 main clusters. First, at the top of Fig. \ref{hclust_m39} we find the first cluster which contains the JIF, SJR and other related measures that express citation normalized per document. Followingly, a second cluster contains the Citation Betweenness Centrality and Pagerank measures that rely on the graph-properties of the citation network. The third cluster contains Total Citation rates, various degree centralities and the H-index that express various distribution parameters of total citation counts. At the bottom of Fig. \ref{hclust_m39}, we find the fourth cluster that contains all usage measures.\\

Table \ref{kmeans} lists the results of a 5 cluster k-means analysis of matrix $R$ that further corroborates the observed clustering in the PCA and hierarchical cluster analysis.\\

\begin{table*}
\caption{\label{kmeans} Results of a k-means cluster analysis of measures.}
\begin{center}
\begin{tabular}{c||l|l}
Cluster	&	Measures										&	Interpretation								\\\hline\hline
1		&	38											&	Journal Use Probability							\\
2		& 	24, 25, 26, 27, 28, 29, 30, 31, 32, 33,  34,  35, 36, 37		&	Usage measures							\\
3		&	1, 2, 3, 4, 5									&	JIF, SJR, Cites per Document measures		\\
4		&	6, 7, 8, 9, 10, 11, 12, 13, 14, 15						&	Total Citation rates and distributions	\\
5		&	16, 17, 18, 19,  20, 21, 22							&	Citation Betweenness and PageRank	\\\hline\hline
\end{tabular}
\end{center}
\end{table*}

The pattern of clusters indicate that some measures express a more distinct aspect of scientific impact and will thus be farther removed from all other measures. Table \ref{measure_loadings} lists the $\bar{\rho}$ values of each measure, defined as the mean Spearman rank-order correlation of a measure to all other 38 measures in $R$. The $\bar{\rho}$ of Citation Half-Life (ID 23) and the Usage Impact Factor (ID 39) fell below the significance threshold of $p<0.05$ for $N=39$, further justifying their removal as outliers. Most $\bar{\rho}$ values range from 0.6 to 0.7 indicating a moderate but significant congruence in the rankings produced by a majority of measures. However, a cluster of five particular measures has low $\bar{\rho}$ values in the range 0.5-0.6. They form a separate, but poorly defined cluster in the lower bottom-left quadrant of Fig. \ref{PCA_m39_map} (ID 1-5: SJR, Immediacy Index, Citation Undirected Weighted Closeness Centrality, Scimago Cites per Doc, and the 2007 JIF), indicating they produce rankings removed from the ``mainstream'' in Fig. \ref{PCA_m39_map}.

\subsection*{Discussion}

\begin{figure}[ht!]
\begin{center}
\includegraphics[width=4in]{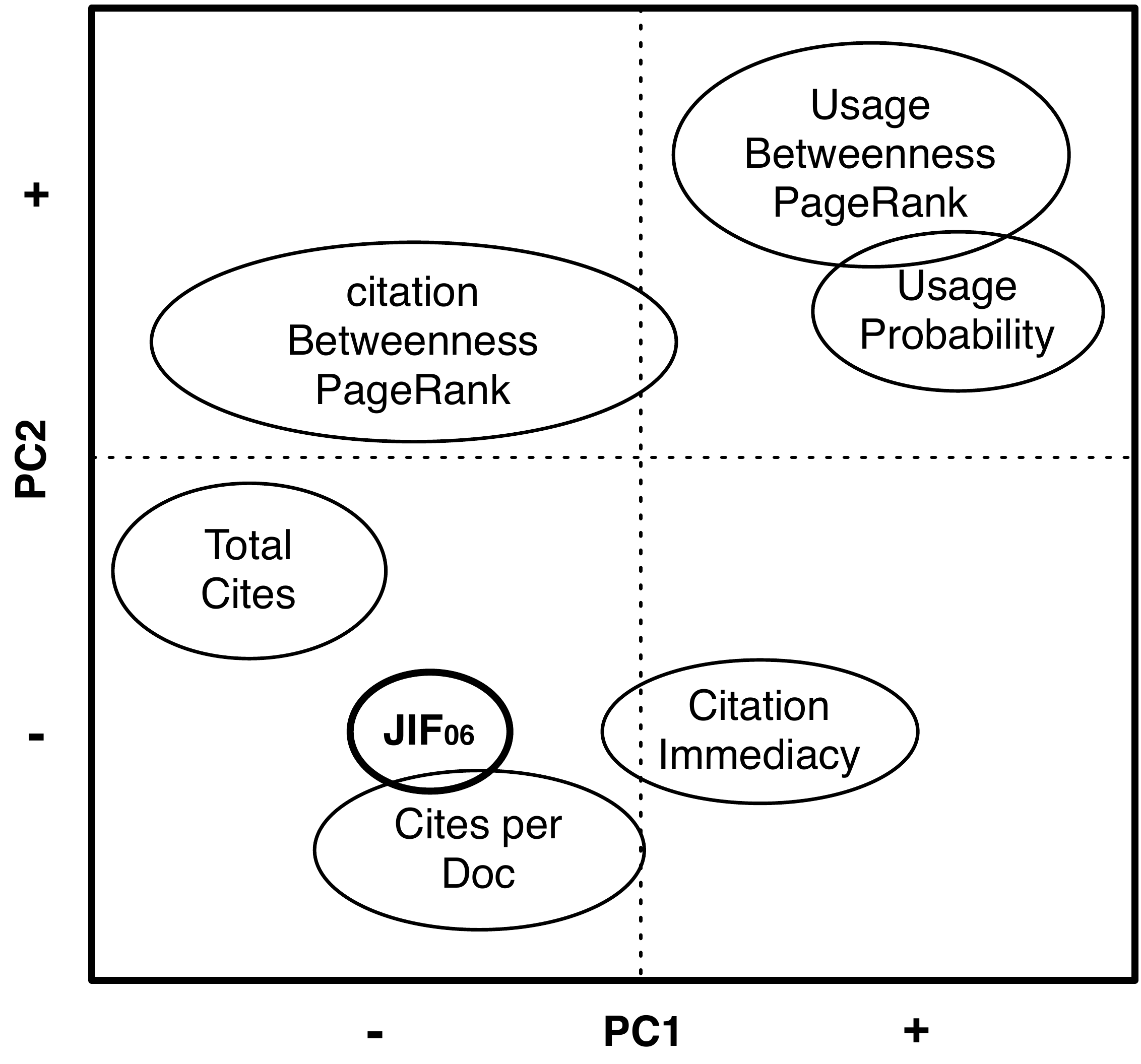}
\caption{\label{PCA_schema}Schematic representation of PCA analysis shown in Fig. \ref{PCA_m39_map}.}
\end{center}
\end{figure}

To interprete the meaning of PC1 and PC2 we need to investigate the distribution of measures along either axis of the map in Fig. \ref{PCA_m39_map}. Fig. \ref{PCA_schema} shows a simplified schema of the distribution of impact measures along the PC1 and PC2 axes. Each of the observed cluster of measures has been given an intuitive ``group'' name to simplify the general pattern.\\

PC1 clearly separates usage measures from citation measures. On the positive end of PC1, we find a sharply demarcated cluster of all usage measures, with the exception of the Journal Use Probability (ID 38) which sits isolated on the extreme positive end of PC1. On the negative end of PC1, we find most citation measures. Surprisingly, some citation measures are positioned close to the cluster of usage measures in terms of their PC1 coordinates. Citation Closeness (ID 3) and in particular Citation Immediacy Index (ID 2) are located on the positive end of PC1, i.e. closest to the usage measures. Citation Betweenness Centrality (IDs 21 and 22) are also positioned closely to the cluster of usage measures according to PC1.\\

This particular distribution of citation measures along PC1 points to an interesting, alternative interpretation of PC1 simply separating the usage from the citation measures. In the center, we find Citation Immediacy Index (ID 2) positioned close to the cluster of usage measures in terms of its PC1 coordinates. The Citation Immediacy Index is intended to be a ``rapid'' indicator of scientific impact since it is based on same-year citations. Its proximity to the usage measures according to PC1 may thus indicate that the usage measures are equally rapid indicators, if not more so.  The assumption that usage measures are "Rapid" indicators of scientific impact is furthermore warranted for the following reasons. First, usage log data is generally considered a more ``rapid'' indicator of scientific impact than citation data, since usage log data is nearly immediately affected by changes in scientific habits and interests whereas citation data is subject to extensive publication delays. It has in fact been shown that present usage rates predict future citation rates \cite{earlie:brody2006}. Second, our usage log data was recorded slightly more recently (April 2006 through March 2007) than the 2007 JCR citation data (January 2006 through December 2007). It may therefore reflect more recent scientific activity. These observations combined lead to a speculative interpretation of PC1 in terms of ``Rapid'' vs. ``Delayed'' measures of impact. The ``Rapid'' indicators are mostly usage measures due to the nature of the usage log data that they have been calculated for. They are however approximated by the Citation Immediacy Index whose definition focuses on same-year citation statistics and two Citation Betweenness Centrality measures (IDs 21 and 22) that may, due to their focus on interdisciplinary power, anticipate emerging scientific activities.\\

PC2 separates citation statistics such as Scimago Total Cites (ID12), JIF (Table \ref{measure_loadings}, ID 5) and Cites per Doc (ID 4) on its negative end from the social network measures such as Citation Betweenness centrality (IDs 21 and 22) and Citation PageRank (ID 16-19) including the Y-factor (ID 20) on its positive end. Measures such as the JIF (ID 5), Scimago Total Cites (ID 12), Journal Cite Probability (ID13), and Journal Use Probability (ID 38) express the rate at which journals indiscriminately receive citations or usage from a variety of sources, i.e.~their Popularity, whereas the mentioned social network measures rely on network structure to express various facets of journal Prestige \cite{journa:bollen2006} or interdisciplinary power \cite{betwee:leydesdorff2007}. PC2 can thus plausibly be interpreted as separating impact measures according to whether they stress scientific Popularity vs. Prestige.\\

Consequently, the PCA results could be interpreted in terms of a separation of measures along two dimensions: ``Rapid'' vs.``Delayed'' (PC1) and ``Popularity'' vs. ``Prestige'' (PC2). Surprisingly, most usage-based measures would then fall in the ``Rapid, ``Prestige'' quadrant, approximated in this aspect only by two Citation Betweenness Centrality measures. The majority of citation-based measures can then be classified as ``Delayed", but with the social network measures being indicative of aspects of ``Prestige'' and the normalized citation measures such as the JIF, Scimago Journal Rank (ID 1) and Cites per Doc  indicative of journal ``Popularity". We also note that the Scimago Journal Rank is positioned among measures such as the JIF and Cites per Doc. This indicates it too expresses ``Delayed'' ``Popularity", in spite of the fact that SJR rankings originate from 2007 citation data and that the SJR has been explicitly defined to ``transfer(s) (of) prestige from a journal to another one" (\url{http://www.scimagojr.com/SCImagoJournalRank.pdf}).\\

Another interesting aspect of the distribution of measures along PC1 and PC2 relates to the determination of a "consensus" view of scientific impact. The $\bar{\rho}$ values indicate the average Spearman rank-order correlation of a particular measure to all other measures, i.e.~the degree to which it approximates the results of all other measures. The measure which best succeeds in approximating the most general sense of scholarly impact will therefore have the highest  $\bar{\rho}$ and will therefore be the best candidate for a "consensus" measure. As shown in Table \ref{measure_loadings} that measure would be Usage Closeness Centrality (ID: 25) whose $\bar{\rho}=0.731$. Conversely, the Citation Scimago Journal Rank (ID1), Citation Immediacy Index (ID 2), Citation Closeness Centrality (ID 3), Citaton Cites per doc (ID 4) and Citation Journal Impact Factor (ID:5) have the lowest $\bar{\rho}$ values indicating that they represent the most particular view of scientific impact.

\subsection*{Future research}

The presented results pertain to what we believe to be the largest and most thorough survey of usage- and citation based measures of scientific impact. Nevertheless, a number of issues need to be addressed in future research efforts.\\

First, although an attempt was made to establish a representative sample of existing and plausible scientific impact measures, several other conceivable impact measures could have been included in this analysis. For example, the HITS algorithm has been successfully applied to web page rankings. Like Google's PageRank it could be calculated for our citation and usage journal networks. Other possible measures that should be considered for inclusion include the Eigenfactor.org measures, and various information-theoretical indexes. The addition of more measures may furthermore enable statistical significance to be achieved on the correlations with now-removed measures such as Citation Half-Life and the Usage Impact Factor, so that they could be included on the generated PCA map of measures.\\

Second, we projected measure correlations onto a space spanned by the 2 highest-ranked components, the first of which seems to make a rather superficial distinction between usage- and citation-derived impact measures and the second of which seems to make a meaningful distinction between "degree" and "quality" of endorsement. Future analysis should focus on including additional components, different combinations of lower-valued components and even the smallest-valued components to determine whether they reveal additional useful distinctions. In addition, non-linear dimensionality reduction methods could be leveraged to reveal non-linear patterns of measure correlations.\\

Third, a significant number of the measures surveyed in this article have been standard tools for decades in social network analysis, but they are not in common use in the domain of scientific impact assessment. To increase the "face-validity" of these rankings, all have been made available to the public on the MESUR web site and can be freely explored and interacted with by users at the following URL: \url{http://www.mesur.org/services}.\\

Fourth, the implemented MESUR services can be enhanced to support the development of novel measures by allowing users to submit their own rankings which can then automatically be placed in the context of existing measures. Such a service could foster the free and open exchange of scientific impact measures by allowing the public to evaluate where any newly proposed measure can be positioned among existing measures. If the measure is deemed to similar to existing measures, it need not be developed. If however, it covers a part of the measure space that was previously unsampled, the new measure may make a significant contribution and could therefore be considered for wider adoption by those involved in scientific assessment.

\subsection*{Conclusion}

Our results indicate that scientific impact is a multi-dimensional construct. The component loadings of a PCA indicate that 92\% of the variances between the correlations of journal rankings produced by 37 impact measures can be explained by the first 3 components. To surpass the 95\% limit, a 4-component model would have to be adopted.\\

A projection of measure correlations onto the first 2 components (83.4\%) nevertheless reveals a number of useful distinctions.  We found that the most salient distinction is made by PC1 which separates usage from citation measures with the exception of Citation Betweenness centrality and Citation Immediacy. The position of the latter and the time periods for which usage was recorded suggests an interpretation of PC1 as making a distinction between measures that provide a "rapid" vs "delayed" view of scientific impact.\\

PC2 seems to separate measures that express Popularity from those that express Prestige. Four general clusters of impact measures can be superimposed on this projection: (1) usage measures, (2) a group of distinctive yet dispersed measures expressing per document citation popularity, (3) measures based on total citation rates and distributions, and (4) finally a set of citation social network measures. These 4 clusters along with the PCA components allows us to quantitatively interpret the landscape of presently available impact measures and determine which aspects of scientific impact they represent. Future research will focus on determining whether these distinctions are stable across a greater variety of measures as well other usage and citation data sets.\\

Four more general conclusions can be drawn from these results; each has significant implications for the developing science of scientific assessment.\\

First, the set of usage measures is more strongly correlated (average Spearman rank-order correlation = 0.93, incl. Usage Probability) than the set of citation measures (average Spearman rank-order correlation = 0.65). This indicates a greater reliability of usage measures calculated from the same usage log data than between citation measures calculated from the same citation data. This effect is possibly caused by the significantly greater density of the usage matrix $U$ in comparison to the citation matrix $C$. As mentioned in the introduction, the amount of usage data that can be collected is much higher than the total amount of citation data in existence because papers can contain only a limited set of citations and once they are published that set is fixed in perpetuity. This limitation may place an upper bound on the reliability that can be achieved with citation measures, but it does not apply to usage measures.\\

Second, if our interpretation of PC2 is correct, usage-based measures are actually {\em stronger} indicators of scientific Prestige than many presently available citation measures. Contrary to expectations, the IF as well as the SJR most strongly express scientific Popularity.\\

Third, some citation measures are more closely related to their usage counterparts than they are to other citation measures such as the JIF.  For example, the Spearman rank-order correlation between Citation Betweenness Centrality and Usage Betweenness Centrality is 0.747. In comparison, the Spearman rank-order correlation between the JIF and Citation Betweenness Centrality is only 0.52. This indicates that contrary to what would be expected, usage impact measures can be closer to a ``consensus ranking'' of journals than some common citation measures.\\

Fourth, and related, when we rank measures according to their average correlation to all other measures $\bar{\rho}$, i.e.~how close they are to all other measures, we find that the JIF and SJR rank 34rd and 38th respectively among 39 measures, indicating their isolated position among the studied set of measures. The JCR Citation Immediacy Index and the Scimago Cites per Doc are in a similar position. On the other hand, Usage Closeness centrality (measure 25) is positioned closest to all other measures (max. $\bar{\rho}=0.731$). These results should give pause to those who consider the JIF the ``golden standard'' of scientific impact. Our results indicate that the JIF and SJR express a rather particular aspect of scientific impact that may not be at the core of the notion of scientific ``impact". Usage-based measures such as Usage Closeness centrality may in fact be better  ``consensus'' measures.


\newpage
\section*{Data files}

The ranking data produced to support the discussed Principal Component Analysis is available  upon request from the corresponding author with the exception of those that have been obtained under proprietary licenses.

\newpage

\bibliographystyle{plos}
\bibliography{PLoSOne_jbollen09.bib}

\end{document}